# Observing nonlinear effects in vibrating strings

Miguel C. Brito[1], Jorge Maia Alves, João Serra, and António Vallera

Departamento Física, CFMC, Faculdade de Ciências Universidade de Lisboa

Edificio C8, Campo Grande, 1749-016 Lisboa, Portugal

**Abstract**

Nonlinear effects were observed in a forced vibrating string. The motion of the string becomes elliptic as the amplitude of the vibration increases. The fundamental resonance frequency depends on the amplitude of the vibration. At sufficient high amplitudes the system exhibits hysterisis: the resonance frequency is different when measured increasing the driving frequency or decreasing it. The experimental results are well understood in the framework of a 2-dimensional model assuming that the string tension is not constant due to the varying length of the string throughout the oscillation.

## 1. Introduction

In the framework of our classes on the experimental study of wave motion we have developed a kit for the study of a vibrating string. A copper wire string is held in mechanical tension between two points (see Fig. 1). An AC current passing through the wire, coupled with the magnetic field of a permanent magnet, produces the mechanical vibration of the string. With this setup we can change the mechanical tension applied to the string, as well as its length and the excitation frequency. The AC current applied to the wire is provided by a specially designed power amplifier with a maximum peak current output of 6A and a bandwidth of 1Hz-1MHz. The possibility of using such high currents enables us to explore the nonlinear behaviour of the vibrating string.

*Fig.1 –The experimental setup: string is held at (1) and (2); AC current generator (3); magnet (4) and weight (5).*

---

[1] mcbrito@fc.ul.pt

It is commonly observed that the motion of the string is transversal (polarized) for small amplitudes but, as the amplitude of the oscillation increases, the string goes into a circular or tubular motion. The symmetry of this circular motion may be broken by assymetric string mounts or if the wire is magnetic (e.g. steel).

When scanning the frequency, there are amplitude discontinuities (jump phenomena) and the resonance frequencies show hysteresis. These nonlienar effects can be understood as arising from the fact that the tension cannot be held constant if the string is varying in length due to the finite amplitude of displacement.

The vibrating string has been shown to have more complex behaviour, such as quasi-periodic and chaotic motions, in particular when damping and excitation are very large [1-3]. These effects however are not discussed in this paper as they are not suitable for an introductory course on experimental study of wave motion.

## 2. Equations of motion

Several authors [1-6] have shown that the equation of motion of the vibrating string can be written in the form

$$\ddot{r} + R\dot{r} + \omega_0^2 r(1 + K r^2) = f(\omega t) \qquad (1)$$

where $r$ is the displacement of an element of the string in the $xy$-plane (which is the plane perpendicular to the length of the string), $K$ is the nonlinear coefficient that takes into account the finite stretching of the string and

$$\omega_0 = k\sqrt{\frac{T_0}{\mu}} \qquad (2)$$

with $T_0$ being the average tension of the string, $\mu$ its linear mass density and $k = 2\pi/\lambda$ where $\lambda$ is the wavelenth. $f(\omega t)$ represents the forcing term and $R$ the damping coefficient. We assume that the ends of the string are symmetrically fixed and thus $R$ is a scalar.

For instance, Tufillaro [1] considered the string as a massless spring with a mass attached at its centre. This mass is subject to damping and forcing. The equation of motion is derived by expanding the restoring force on the mass in a series, whilst neglecting fourth and higher power terms. In Gough [4] and Elliot [5] the constant tension in the linear equation is replaced by its Hooke's law expression. With this approach they were able to relate the parameter $K$ with actual string parameters. More recently Bolwell [6] rewrote the equation of motion as

$$\ddot{r} + R\dot{r} + r\left(c_1^2 + \frac{1}{2}c_2^2 r^2\right) = f(\omega t) \tag{3}$$

where he identified the parameter $c_1$ as the transverse wave velocity and

$$c_2 = k\sqrt{\frac{YA}{\mu}} \tag{4}$$

as the compressional longitudinal wave velocity. $Y$ represents the string Young modulus and $A$ its cross section.

If we neglect the tension oscillation due to the elongation of the string ($K = 0$ or $c_2 = 0$), the compressional longitudinal wave vanishes and we recover the usual linear equation of motion of a vibrating string. Assuming that the string is being sinusoidally excited in the $x$ direction, the forcing term becomes $F\cos(\omega t)$. Furthermore, writing the equation of motion explicitly we get

$$\ddot{x} + R(\dot{x} + \dot{y}) + x\left[c_1^2 + \frac{1}{2}c_2^2(x^2 + y^2)\right] = F\cos(\omega t)$$

$$\ddot{y} + R(\dot{x} + \dot{y}) + y\left[c_1^2 + \frac{1}{2}c_2^2(x^2 + y^2)\right] = 0 \tag{5}$$

These equations describe two parametrically coupled oscillators with cubic nonlinearities. Exciting a nonlinear resonance in one oscillator causes the parametric excitation of linear oscillation in the second, which then couples back into the first. One should also notice that the nonlinearity is stronger for higher values of $c_2$, i.e. when the effect of the longitudinal wave becomes important.

## 3. Solutions of the equations of motion

In order to gain some insight on the physics of the vibrating string, we shall start by looking at the undamped free oscillation discussing both the circular and the general solutions. We shall be concentrating on the single mode solution, although one of the effects of the nonlinearity could be the excitation of higher order modes. Then, we shall determine an approximate solution for the equations of motion (5) for the damped forced solution. Finally, we calculate the frequency-response curve for circular motion and discuss the results.

### 3.1. Circular motion for undamped free oscillation

The simplest solution to the equations of motion for undamped free oscillation equation $(R = F = 0)$ is circular motion with constant radius, $A$,

$$x = A\cos(\omega t)$$
$$y = A\sin(\omega t) \qquad (6)$$

In this situation the elastic restoring force acts as a centripetal force responsible for the circular motion. Introducing these solutions into Eq. (5) we get

$$\omega^2 = c_1^2 + \frac{1}{2}c_2^2 A^2 \qquad (7)$$

which means that there is an adjustment to the motion frequency due to the stretching of the string. As expected, we can see that this de-tuning of the resonance frequency is larger for increasing amplitudes.

### 3.2. General solution for undamped free oscillation

The general solution for the undamped free vibrating string equations, determined by considering a reference frame rotating at angular velocity $\Omega$, can be shown to be [5]

$$x = \frac{A+B}{2}\cos((c_1 + \Omega)t) + \frac{A-B}{2}\cos((c_1 - \Omega)t)$$
$$y = \frac{A+B}{2}\sin((c_1 + \Omega)t) + \frac{A-B}{2}\sin((c_1 - \Omega)t) \qquad (8)$$

We can see that the general solution is a linear superposition of two circular motions, of opposite senses, at frequencies $c_1 + \Omega$ and $c_1 - \Omega$. The precession frequency $\Omega$ is given by

$$\Omega = c_2 \frac{AB}{2} \qquad (9)$$

Once more, the effect of precession is stronger for larger amplitudes.

### 3.3. Approximate solution for damped forced oscillation

For the damped forced oscillations of the vibrating string we start by assuming that the motion will take place at the driving frequency and is composed of two waves travelling in opposite directions:

$$x = \frac{1}{2}\left(A e^{i\omega t} + A^* e^{-i\omega t}\right)$$

$$y = \frac{1}{2}\left(B e^{i\omega t} + B^* e^{-i\omega t}\right) \quad (10)$$

Introducing these solutions into the equations of motion (5) and ignoring all terms not at the driving frequency we get the following equations

$$A\left(c_1^2 + iR\omega - \omega^2\right) + \frac{1}{2}c_2^2\left(\frac{3}{4}A^2 A^* + \frac{1}{4}B^2 A^* + \frac{1}{2}ABB^*\right) = 2F$$

$$B\left(c_1^2 + iR\omega - \omega^2\right) + \frac{1}{2}c_2^2\left(\frac{3}{4}B^2 B^* + \frac{1}{4}A^2 B^* + \frac{1}{2}BAA^*\right) = 0 \quad (11)$$

Writing the complex amplitudes $A$ and $B$ in the form $A = a e^{i\alpha}$ and $B = b e^{i\beta}$ (where $a$, $b$, $\alpha$ and $\beta$ are all real constants) we get, from the imaginary part of the previous equations,

$$\frac{1}{8}c_2^2\left(a^2 + b^2\right)\sin[2(\alpha - \beta)] = 0 \quad \Rightarrow \quad \alpha = \beta + n\frac{\pi}{2} \quad (12)$$

with $n$ being an integer. From the real part of the equations, for even $n$, we get

$$c_1^2 + iR\omega - \omega^2 + \frac{1}{2}c_2^2\left(\frac{3}{4}a^2 + \frac{3}{4}b^2\right) = \frac{2F}{a}$$

$$c_1^2 + iR\omega - \omega^2 + \frac{1}{2}c_2^2\left(\frac{3}{4}b^2 + \frac{3}{4}a^2\right) = 0 \quad (13)$$

which has no physical solution. For odd $n$, we get

$$c_1^2 + iR\omega - \omega^2 + \frac{1}{2}c_2^2\left(\frac{3}{4}a^2 + \frac{1}{4}b^2\right) = \frac{2F}{a}$$

$$c_1^2 + iR\omega - \omega^2 + \frac{1}{2}c_2^2\left(\frac{3}{4}b^2 + \frac{1}{4}a^2\right) = 0 \quad (14)$$

that can be combined to yield the trajectory traced by the string in the $xy$-plane:

$$ab^2 - a^3 = \frac{8F}{c_2^2} \qquad (15)$$

This result agrees with the solution determined by Elliot [7] using the Duffing method. The solution to Eq. (15) is plotted in Fig. 2. For small amplitudes, bellow a critical amplitude $a_{crit}$ given by

$$a_{crit} = \sqrt[3]{\frac{8F}{c_2^2}} \qquad (16)$$

$b$ is imaginary and has no real solution, and the motion becomes a purely polarized oscillation in the driving plane. This is the solution of the linear equation in which there is no coupling between the motions in the driving $x$-direction and the $y$-direction.

As the amplitude increases pass $a_{crit}$, $b$ increases towards $b = a$ and the motion becomes elliptic. This is due to the strong tension increase with increasing amplitude motion. The phase between the motions in the two directions is then an odd multiple of $\pi/2$.

Fig.2 - Solution of Eq. (15): the amplitude of the string along the x and the y directions.

### 3.4. Frequency-response curve

Experimentally, the easiest procedure to test this model of the vibrating string is to measure the frequency-response curve, i.e. the amplitude of the motion as a function of the driving frequency.

When the string is vibrating near the resonance the amplitude of the oscillation becomes large and, according to the results of the previous section, the motion of the string becomes asymptotically circular. Imposing equal amplitudes and a phase shift of $n\pi/2$ to solutions of the from (10), the equation of motion (5) of the damped forced string becomes

$$\ddot{x} + x(c_1^2 + c_2^2 x^2) + 2R\dot{x} = F\cos(\omega t) \qquad (17)$$

with

$$x = \frac{1}{2}\left(Ae^{i\omega t} + A^* e^{-i\omega t}\right) \qquad (18)$$

Introducing this into the equation of motion (17) and ignoring all terms not at the driving frequency we get

$$\left( c_1^2 - \omega^2 + i2\omega R + \frac{3}{2} c_2^2 |A|^2 \right) A = F \quad (19)$$

Defining the real numbers $a$ and $\alpha$ such that $A = a e^{i\alpha}$, we can rewrite Eq. (19) as

$$\begin{cases} 2R\omega a = F \sin\alpha \\ \left[ \left( c_1^2 - \omega^2 + \frac{3}{2} c_2^2 a^2 \right) \right] a = F \cos\alpha \end{cases} \quad (20)$$

Squaring and adding the two equations we obtain

$$\left[ (2R\omega)^2 + \left( c_1^2 - \omega^2 + \frac{3}{2} c_2^2 a^2 \right)^2 \right] a^2 = F \quad (21)$$

that describes the frequency-response curve of the vibrating string (see Fig. 3).

*Fig. 3 - a) Frequency-response curve of a vibrating string with (solid line) and without (dashed line) nonlinear term; b) Frequency-response curve of nonlinear vibrating string with (solid line) and without (dashed line) damping term.*

The most striking feature of this plot is the bending of the resonance due to the nonlinearity introduced by the variation of the tension. This result indicates, and experiment confirms, that as the frequency is slowly increased the resonance frequency shifts towards higher values. When we get over the resonance, the system "jumps" down to the lowest amplitude available for that frequency, and the circular motion collapses.
The frequency-response curve also shows that the vibrating string exhibits hysteresis: the resonance frequency is different when measured increasing the driving frequency or decreasing it.

## 4. Experimental measurements

### 4.1. Resonant frequency

Figure 4 shows a typical measurement of the frequency-response curve for a 0.5m long copper string ($\mu = 6.9 \ 10^{-4}$ kg m$^{-1}$, $Y = 1.2 \ 10^{11}$ Pa) with 0.3mm diameter, and its fit using equation 14. For the fit, the compression longitudinal wave velocity $c_2$ was assumed known (using Eq. 4) and the free parameters were $F$, $R$ and $c_1$. The measured tension was 3.0N and the force needed to lateraly displace the middle of the string 10mm away from its rest position was 0.2N.

*Fig. 4 - Fit to the experimentally measured frequency-response curve for the fundamental harmonic mode of a 0.5m long copper string with 0.3mm diameter.*

The fit is well adjusted to the experimental data. We can compare the adjusted parameters with the actual parameters of the string. The driving force $F$ is compatible with its experimental measurement. From $c_1$, and using Eq. (4), one determines the average tension ($T_0 = 3.2$N), which is also rather close to the tension measured experimentally.

### 4.2. String polarisation

The polarization of the string, and its dependence on the amplitude, was observed by improving the optical contrast of the motion of the string with a painted white dot on the centre of the string. The dot is illuminated with a desk lamp.
Inserting the parameters adjusted in the previous section into Eq. (15), one gets the trajectory of the string in the *xy*-plane. As discussed in section 3.3, for small amplitudes the motion is purely polarized in the plane of the driving force. For amplitudes larger than $a_{crit} = 3$mm the motion of the string becomes elliptic.
This result was confirmed experimentally.

## 5. Conclusions

The nonlinear behviour of the vibrating string can be understood as arising from the fact that the tension is not constant due to the varying length of the string throughout the oscillation. This effect leads to the propagation of compressional longitudinal waves.

We have developed a 2-dimensional model where, for each plane, the motion takes place at the driving frequency and is composed of two waves travelling in opposite directions. The phase difference between the motion in the driving plane and the transverse plane was shown to be an odd multiple of π/2.

We have shown that the motion of the string is restricted to the driving plane for small amplitudes and above an amplitude of $a_{crit}$ the motion becomes elliptic. This effect was observed experimentally. We can then say that the linear equation of motion of the vibrating string is thus valid for amplitudes smaller than $a_{crit}$.

Furthermore, the study of the frequency-response curve has shown that the jump phenomena and hysterisis are reproduced by the model and the fit to the experimental data is well adjusted to the actual parameters of the string.

# CAPTIONS

Fig.1 –The experimental setup: string is held at (1) and (2); AC current generator (3); magnet (4) and weight (5).

Fig.2 - Solution of Eq. (15): the amplitude of the string along the *x* and the *y* directions.

Fig. 3 - a) Frequency-response curve of a vibrating string with (solid line) and without (dashed line) nonlinear term; b) Frequency-response curve of nonlinear vibrating string with (solid line) and without (dashed line) damping term.

Fig. 4 - Fit to the experimentally measured frequency-response curve for the fundamental harmonic mode of a 0.5m long copper string with 0.3mm diameter.

# FIGURES

Figure 1:

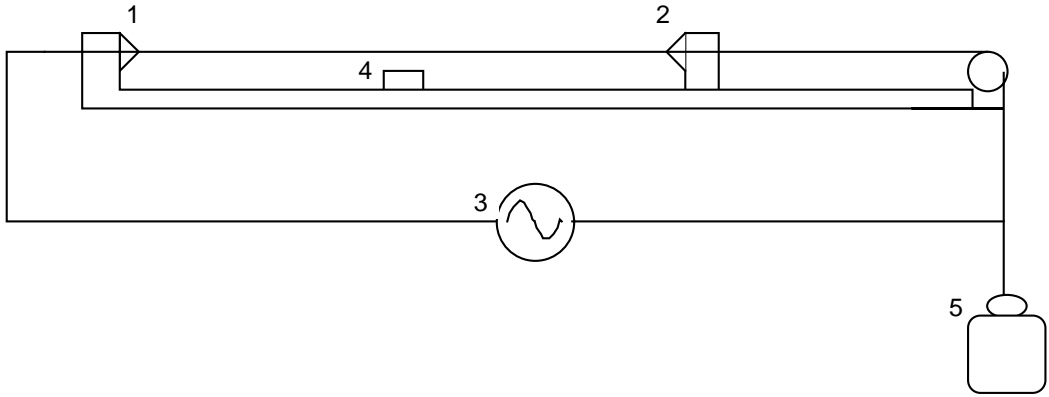

Figure 2:

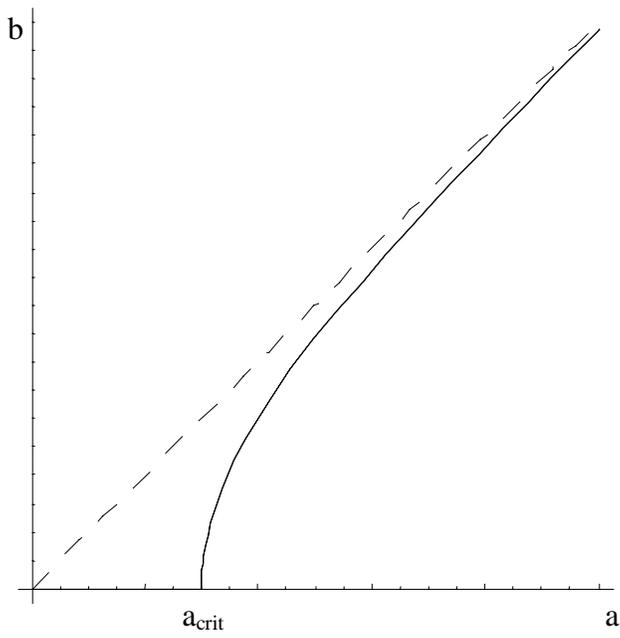

Figure 3:

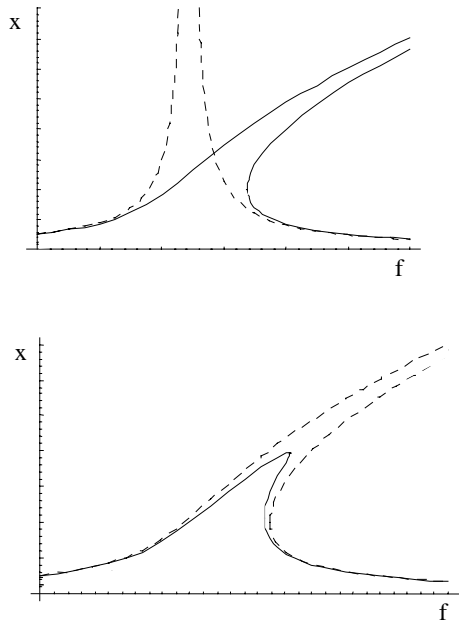

Figure 4:

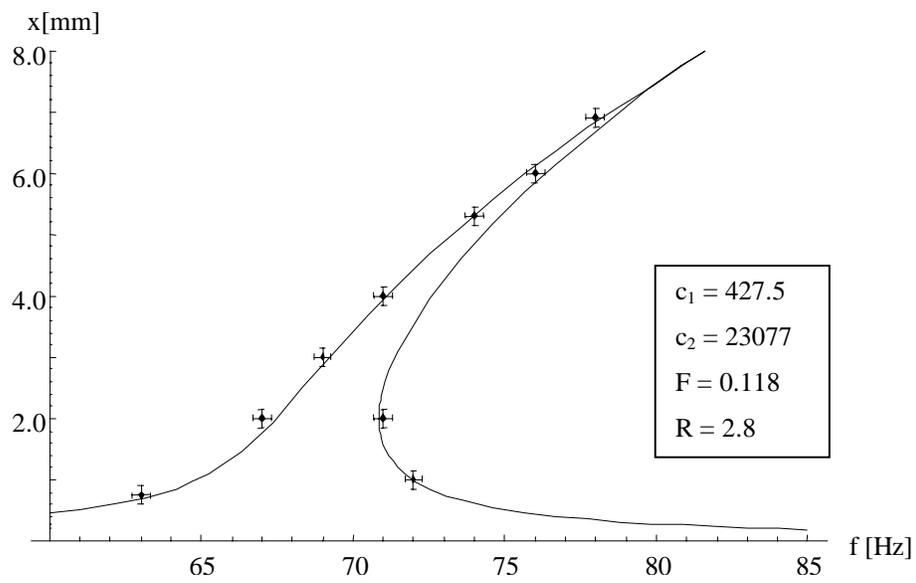